\documentclass[final]{IEEEtran}
\usepackage[update,prepend]{epstopdf}

\usepackage{graphics}
\usepackage{multirow}
\usepackage{tikz}
\usepackage{bbm} 
\usepackage{pdfpages}
\usepackage{multirow}
\usepackage{subfig}
\usepackage{comment}
\usepackage{makecell}
\usepackage{epsfig}
\usepackage{epstopdf}
\usepackage{caption} 
\usepackage{setspace}	
\usepackage{graphicx}
\usepackage{algorithm,algorithmic}
\usepackage{multicol}

\usepackage[justification=centering]{caption}
\usepackage{textcomp}
\usepackage{psfrag}
\usepackage{arydshln}
\usepackage{url}
\usepackage{soul}
\usepackage{graphicx,color}
\usepackage[nolist]{acronym}
\usepackage{array}

\usepackage{mathtools,lipsum}
\usepackage{cuted}
\usepackage{amsmath}
\usepackage{graphicx}
\usepackage{threeparttable}

\def\nb0{{\mathbf{0}}}
\def\nb1{{\mathbf{1}}}

\begin{document}
\graphicspath{{./Figures/}}
	\begin{acronym}

\acro{5G-NR}{5G New Radio}
\acro{3GPP}{3rd Generation Partnership Project}
\acro{ABS}{aerial base station}
\acro{AC}{address coding}
\acro{ACF}{autocorrelation function}
\acro{ACR}{autocorrelation receiver}
\acro{ADC}{analog-to-digital converter}
\acrodef{aic}[AIC]{Analog-to-Information Converter}     
\acro{AIC}[AIC]{Akaike information criterion}
\acro{aric}[ARIC]{asymmetric restricted isometry constant}
\acro{arip}[ARIP]{asymmetric restricted isometry property}

\acro{ARQ}{Automatic Repeat Request}
\acro{AUB}{asymptotic union bound}
\acrodef{awgn}[AWGN]{Additive White Gaussian Noise}     
\acro{AWGN}{additive white Gaussian noise}

\acro{APSK}[PSK]{asymmetric PSK} 

\acro{waric}[AWRICs]{asymmetric weak restricted isometry constants}
\acro{warip}[AWRIP]{asymmetric weak restricted isometry property}
\acro{BCH}{Bose, Chaudhuri, and Hocquenghem}        
\acro{BCHC}[BCHSC]{BCH based source coding}
\acro{BEP}{bit error probability}
\acro{BFC}{block fading channel}
\acro{BG}[BG]{Bernoulli-Gaussian}
\acro{BGG}{Bernoulli-Generalized Gaussian}
\acro{BPAM}{binary pulse amplitude modulation}
\acro{BPDN}{Basis Pursuit Denoising}
\acro{BPPM}{binary pulse position modulation}
\acro{BPSK}{Binary Phase Shift Keying}
\acro{BPZF}{bandpass zonal filter}
\acro{BSC}{binary symmetric channels}              
\acro{BU}[BU]{Bernoulli-uniform}
\acro{BER}{bit error rate}
\acro{BS}{base station}
\acro{BW}{BandWidth}
\acro{BLLL}{ binary log-linear learning }

\acro{CP}{Cyclic Prefix}
\acrodef{cdf}[CDF]{cumulative distribution function}   
\acro{CDF}{Cumulative Distribution Function}
\acrodef{c.d.f.}[CDF]{cumulative distribution function}
\acro{CCDF}{complementary cumulative distribution function}
\acrodef{ccdf}[CCDF]{complementary CDF}               
\acrodef{c.c.d.f.}[CCDF]{complementary cumulative distribution function}
\acro{CD}{cooperative diversity}

\acro{CDMA}{Code Division Multiple Access}
\acro{ch.f.}{characteristic function}
\acro{CIR}{channel impulse response}
\acro{cosamp}[CoSaMP]{compressive sampling matching pursuit}
\acro{CR}{cognitive radio}
\acro{cs}[CS]{compressed sensing}                   
\acrodef{cscapital}[CS]{Compressed sensing} 
\acrodef{CS}[CS]{compressed sensing}
\acro{CSI}{channel state information}
\acro{CCSDS}{consultative committee for space data systems}
\acro{CC}{convolutional coding}
\acro{Covid19}[COVID-19]{Coronavirus disease}

\acro{DAA}{detect and avoid}
\acro{DAB}{digital audio broadcasting}
\acro{DCT}{discrete cosine transform}
\acro{dft}[DFT]{discrete Fourier transform}
\acro{DR}{distortion-rate}
\acro{DS}{direct sequence}
\acro{DS-SS}{direct-sequence spread-spectrum}
\acro{DTR}{differential transmitted-reference}
\acro{DVB-H}{digital video broadcasting\,--\,handheld}
\acro{DVB-T}{digital video broadcasting\,--\,terrestrial}
\acro{DL}{DownLink}
\acro{DSSS}{Direct Sequence Spread Spectrum}
\acro{DFT-s-OFDM}{Discrete Fourier Transform-spread-Orthogonal Frequency Division Multiplexing}
\acro{DAS}{Distributed Antenna System}
\acro{DNA}{DeoxyriboNucleic Acid}

\acro{EC}{European Commission}
\acro{EED}[EED]{exact eigenvalues distribution}
\acro{EIRP}{Equivalent Isotropically Radiated Power}
\acro{ELP}{equivalent low-pass}
\acro{eMBB}{Enhanced Mobile Broadband}
\acro{EMF}{ElectroMagnetic Field}
\acro{EU}{European union}
\acro{EI}{Exposure Index}
\acro{eICIC}{enhanced Inter-Cell Interference Coordination}

\acro{FC}[FC]{fusion center}
\acro{FCC}{Federal Communications Commission}
\acro{FEC}{forward error correction}
\acro{FFT}{fast Fourier transform}
\acro{FH}{frequency-hopping}
\acro{FH-SS}{frequency-hopping spread-spectrum}
\acrodef{FS}{Frame synchronization}
\acro{FSsmall}[FS]{frame synchronization}  
\acro{FDMA}{Frequency Division Multiple Access}

\acro{GA}{Gaussian approximation}
\acro{GF}{Galois field }
\acro{GG}{Generalized-Gaussian}
\acro{GIC}[GIC]{generalized information criterion}
\acro{GLRT}{generalized likelihood ratio test}
\acro{GPS}{Global Positioning System}
\acro{GMSK}{Gaussian Minimum Shift Keying}
\acro{GSMA}{Global System for Mobile communications Association}
\acro{GS}{ground station}
\acro{GMG}{ Grid-connected MicroGeneration}

\acro{HAP}{high altitude platform}
\acro{HetNet}{Heterogeneous network}

\acro{IDR}{information distortion-rate}
\acro{IFFT}{inverse fast Fourier transform}
\acro{iht}[IHT]{iterative hard thresholding}
\acro{i.i.d.}{independent, identically distributed}
\acro{IoT}{Internet of Things}                      
\acro{IR}{impulse radio}
\acro{lric}[LRIC]{lower restricted isometry constant}
\acro{lrict}[LRICt]{lower restricted isometry constant threshold}
\acro{ISI}{intersymbol interference}
\acro{ITU}{International Telecommunication Union}
\acro{ICNIRP}{International Commission on Non-Ionizing Radiation Protection}
\acro{IEEE}{Institute of Electrical and Electronics Engineers}
\acro{ICES}{IEEE international committee on electromagnetic safety}
\acro{IEC}{International Electrotechnical Commission}
\acro{IARC}{International Agency on Research on Cancer}
\acro{IS-95}{Interim Standard 95}

\acro{KPI}{Key Performance Indicator}

\acro{LEO}{low earth orbit}
\acro{LF}{likelihood function}
\acro{LLF}{log-likelihood function}
\acro{LLR}{log-likelihood ratio}
\acro{LLRT}{log-likelihood ratio test}
\acro{LoS}{Line-of-Sight}
\acro{LRT}{likelihood ratio test}
\acro{wlric}[LWRIC]{lower weak restricted isometry constant}
\acro{wlrict}[LWRICt]{LWRIC threshold}
\acro{LPWAN}{Low Power Wide Area Network}
\acro{LoRaWAN}{Low power long Range Wide Area Network}
\acro{NLoS}{Non-Line-of-Sight}
\acro{LiFi}[Li-Fi]{light-fidelity}
 \acro{LED}{light emitting diode}
 \acro{LABS}{LoS transmission with each ABS}
 \acro{NLABS}{NLoS transmission with each ABS}

\acro{MB}{multiband}
\acro{MC}{macro cell}
\acro{MDS}{mixed distributed source}
\acro{MF}{matched filter}
\acro{m.g.f.}{moment generating function}
\acro{MI}{mutual information}
\acro{MIMO}{Multiple-Input Multiple-Output}
\acro{MISO}{multiple-input single-output}
\acrodef{maxs}[MJSO]{maximum joint support cardinality}                       
\acro{ML}[ML]{maximum likelihood}
\acro{MMSE}{minimum mean-square error}
\acro{MMV}{multiple measurement vectors}
\acrodef{MOS}{model order selection}
\acro{M-PSK}[${M}$-PSK]{$M$-ary phase shift keying}                       
\acro{M-APSK}[${M}$-PSK]{$M$-ary asymmetric PSK} 
\acro{MP}{ multi-period}
\acro{MINLP}{mixed integer non-linear programming}

\acro{M-QAM}[$M$-QAM]{$M$-ary quadrature amplitude modulation}
\acro{MRC}{maximal ratio combiner}                  
\acro{maxs}[MSO]{maximum sparsity order}                                      
\acro{M2M}{Machine-to-Machine}                                                
\acro{MUI}{multi-user interference}
\acro{mMTC}{massive Machine Type Communications}      
\acro{mm-Wave}{millimeter-wave}
\acro{MP}{mobile phone}
\acro{MPE}{maximum permissible exposure}
\acro{MAC}{media access control}
\acro{NB}{narrowband}
\acro{NBI}{narrowband interference}
\acro{NLA}{nonlinear sparse approximation}
\acro{NLOS}{Non-Line of Sight}
\acro{NTIA}{National Telecommunications and Information Administration}
\acro{NTP}{National Toxicology Program}
\acro{NHS}{National Health Service}

\acro{LOS}{Line of Sight}

\acro{OC}{optimum combining}                             
\acro{OC}{optimum combining}
\acro{ODE}{operational distortion-energy}
\acro{ODR}{operational distortion-rate}
\acro{OFDM}{Orthogonal Frequency-Division Multiplexing}
\acro{omp}[OMP]{orthogonal matching pursuit}
\acro{OSMP}[OSMP]{orthogonal subspace matching pursuit}
\acro{OQAM}{offset quadrature amplitude modulation}
\acro{OQPSK}{offset QPSK}
\acro{OFDMA}{Orthogonal Frequency-division Multiple Access}
\acro{OPEX}{Operating Expenditures}
\acro{OQPSK/PM}{OQPSK with phase modulation}

\acro{PAM}{pulse amplitude modulation}
\acro{PAR}{peak-to-average ratio}
\acrodef{pdf}[PDF]{probability density function}                      
\acro{PDF}{probability density function}
\acrodef{p.d.f.}[PDF]{probability distribution function}
\acro{PDP}{power dispersion profile}
\acro{PMF}{probability mass function}                             
\acrodef{p.m.f.}[PMF]{probability mass function}
\acro{PN}{pseudo-noise}
\acro{PPM}{pulse position modulation}
\acro{PRake}{Partial Rake}
\acro{PSD}{power spectral density}
\acro{PSEP}{pairwise synchronization error probability}
\acro{PSK}{phase shift keying}
\acro{PD}{power density}
\acro{8-PSK}[$8$-PSK]{$8$-phase shift keying}
\acro{PPP}{Poisson point process}
\acro{PCP}{Poisson cluster process}
 
\acro{FSK}{Frequency Shift Keying}

\acro{QAM}{Quadrature Amplitude Modulation}
\acro{QPSK}{Quadrature Phase Shift Keying}
\acro{OQPSK/PM}{OQPSK with phase modulator }

\acro{RD}[RD]{raw data}
\acro{RDL}{"random data limit"}
\acro{ric}[RIC]{restricted isometry constant}
\acro{rict}[RICt]{restricted isometry constant threshold}
\acro{rip}[RIP]{restricted isometry property}
\acro{ROC}{receiver operating characteristic}
\acro{rq}[RQ]{Raleigh quotient}
\acro{RS}[RS]{Reed-Solomon}
\acro{RSC}[RSSC]{RS based source coding}
\acro{r.v.}{random variable}                               
\acro{R.V.}{random vector}
\acro{RMS}{root mean square}
\acro{RFR}{radiofrequency radiation}
\acro{RIS}{Reconfigurable Intelligent Surface}
\acro{RNA}{RiboNucleic Acid}
\acro{RRM}{Radio Resource Management}
\acro{RUE}{reference user equipments}
\acro{RAT}{radio access technology}
\acro{RB}{resource block}

\acro{SA}[SA-Music]{subspace-augmented MUSIC with OSMP}
\acro{SC}{small cell}
\acro{SCBSES}[SCBSES]{Source Compression Based Syndrome Encoding Scheme}
\acro{SCM}{sample covariance matrix}
\acro{SEP}{symbol error probability}
\acro{SG}[SG]{sparse-land Gaussian model}
\acro{SIMO}{single-input multiple-output}
\acro{SINR}{signal-to-interference plus noise ratio}
\acro{SIR}{signal-to-interference ratio}
\acro{SISO}{Single-Input Single-Output}
\acro{SMV}{single measurement vector}
\acro{SNR}[\textrm{SNR}]{signal-to-noise ratio} 
\acro{sp}[SP]{subspace pursuit}
\acro{SS}{spread spectrum}
\acro{SW}{sync word}
\acro{SAR}{specific absorption rate}
\acro{SSB}{synchronization signal block}
\acro{SR}{shrink and realign}

\acro{tUAV}{tethered Unmanned Aerial Vehicle}
\acro{TBS}{terrestrial base station}

\acro{uUAV}{untethered Unmanned Aerial Vehicle}
\acro{PDF}{probability density functions}

\acro{PL}{path-loss}

\acro{TH}{time-hopping}
\acro{ToA}{time-of-arrival}
\acro{TR}{transmitted-reference}
\acro{TW}{Tracy-Widom}
\acro{TWDT}{TW Distribution Tail}
\acro{TCM}{trellis coded modulation}
\acro{TDD}{Time-Division Duplexing}
\acro{TDMA}{Time Division Multiple Access}
\acro{Tx}{average transmit}

\acro{UAV}{Unmanned Aerial Vehicle}
\acro{uric}[URIC]{upper restricted isometry constant}
\acro{urict}[URICt]{upper restricted isometry constant threshold}
\acro{UWB}{ultrawide band}
\acro{UWBcap}[UWB]{Ultrawide band}   
\acro{URLLC}{Ultra Reliable Low Latency Communications}
         
\acro{wuric}[UWRIC]{upper weak restricted isometry constant}
\acro{wurict}[UWRICt]{UWRIC threshold}                
\acro{UE}{User Equipment}
\acro{UL}{UpLink}

\acro{WiM}[WiM]{weigh-in-motion}
\acro{WLAN}{wireless local area network}
\acro{wm}[WM]{Wishart matrix}                               
\acroplural{wm}[WM]{Wishart matrices}
\acro{WMAN}{wireless metropolitan area network}
\acro{WPAN}{wireless personal area network}
\acro{wric}[WRIC]{weak restricted isometry constant}
\acro{wrict}[WRICt]{weak restricted isometry constant thresholds}
\acro{wrip}[WRIP]{weak restricted isometry property}
\acro{WSN}{wireless sensor network}                        
\acro{WSS}{Wide-Sense Stationary}
\acro{WHO}{World Health Organization}
\acro{Wi-Fi}{Wireless Fidelity}

\acro{sss}[SpaSoSEnc]{sparse source syndrome encoding}

\acro{VLC}{Visible Light Communication}
\acro{VPN}{Virtual Private Network} 
\acro{RF}{Radio Frequency}
\acro{FSO}{Free Space Optics}
\acro{IoST}{Internet of Space Things}

\acro{GSM}{Global System for Mobile Communications}
\acro{2G}{Second-generation cellular network}
\acro{3G}{Third-generation cellular network}
\acro{4G}{Fourth-generation cellular network}
\acro{5G}{Fifth-generation cellular network}	
\acro{gNB}{next-generation Node-B Base Station}
\acro{NR}{New Radio}
\acro{UMTS}{Universal Mobile Telecommunications Service}
\acro{LTE}{Long Term Evolution}

\acro{QoS}{Quality of Service}
\end{acronym}
	
\newcommand{\SAR} {\mathrm{SAR}}
\newcommand{\WBSAR} {\mathrm{SAR}_{\mathsf{WB}}}
\newcommand{\gSAR} {\mathrm{SAR}_{10\si{\gram}}}
\newcommand{\Sab} {S_{\mathsf{ab}}}
\newcommand{\Eavg} {E_{\mathsf{avg}}}
\newcommand{\ft}{f_{\textsf{th}}}
\newcommand{\alphatf}{\alpha_{24}}

\title{
High-Altitude Platforms in the Low-Altitude Economy: Bridging Communication, Computing, and Regulation
}
\author{
 
Bang Huang, {\em Member, IEEE}, Baha Eddine Youcef Belmekki, and Mohamed-Slim Alouini, {\em Fellow, IEEE}

\thanks{Bang Huang and Mohamed-Slim Alouini are with King Abdullah University of Science and Technology (KAUST), CEMSE division, Thuwal 23955-6900, Saudi Arabia (e-mail:  bang.huang@kaust.edu.sa; slim.alouini@kaust.edu.sa).}
\thanks{Baha Eddine Youcef Belmekki is  with Heriot-Watt University in Edinburgh, United Kingdom (e-mail:  B.Belmekki@hw.ac.uk).}
\vspace{-6mm}
}
\maketitle
\thispagestyle{empty}

\begin{abstract}
The Low-Altitude Economy (LAE) is rapidly emerging as a new technological and industrial frontier, with unmanned aerial vehicles (UAVs), electric vertical takeoff and landing (eVTOL) aircraft, and aerial swarms increasingly deployed in logistics, infrastructure inspection, security, and emergency response. However, the large-scale development of the LAE demands a reliable aerial foundation that ensures not only real-time connectivity and computational support, but also navigation integrity and safe airspace management for safety-critical operations.
High-Altitude Platforms (HAPs), positioned at around 20 km, provide a unique balance between wide-area coverage and low-latency responsiveness. Compared with low earth orbit (LEO) satellites, HAPs are closer to end users and thus capable of delivering millisecond-level connectivity, fine-grained regulatory oversight, and powerful onboard computing and caching resources. Beyond connectivity and computation, HAPs-assisted sensing and regulation further enable navigation integrity and airspace trust, which are essential for safety-critical UAV and eVTOL operations in the LAE. This article proposes a five-stage evolutionary roadmap for HAPs in the LAE: from serving as aerial infrastructure bases, to becoming super back-ends for UAV, to acting as frontline support for ground users, further enabling swarm-scale UAV coordination, and ultimately advancing toward edge–air–cloud closed-loop autonomy.  
In parallel, HAPs complement LEO satellites and cloud infrastructures to form a global–regional–local three-tier architecture. Looking forward, HAPs are expected to evolve from simple platforms into intelligent hubs, emerging as pivotal nodes for air traffic management, intelligent logistics, and emergency response. By doing so, they will accelerate the transition of the LAE toward large-scale deployment, autonomy, and sustainable growth.
\end{abstract}

\section{Introduction}
\label{sec_introduction}
The Low-Altitude Economy (LAE) is emerging as a promising growth engine following the digital economy. Operating primarily within airspace below 1km and driven by drones, electric vertical takeoff and landing (eVTOL) \cite{zaid2023evtol}, and other low-altitude aerial vehicles, LAE spans diverse application domains, including logistics and delivery, emergency response, smart cities, precision agriculture, and tourism \cite{SongLin2025TrustworthyIntelligent}. The global drones market was valued at USD 73 billion in 2024 and is projected to grow at a compound annual growth rate (CAGR) of 14.3$\%$ from 2024 to 2030, highlighting the strong momentum of the sector\footnote{https://www.grandviewresearch.com/industry-analysis/drone-market-report}. It not only accelerates the commercialization of airspace resources but also fosters a complete industrial value chain, from platform manufacturing and operational services to data platforms. Unsurprisingly, LAE has attracted strong attention from both industry and government. Aerospace leaders such as Airbus\footnote{https://www.airbus.com/en/innovation/future-aircraft-operations/airbus-unmanned-traffic-management} and Boeing\footnote{https://www.boeing.com/innovation/innovation-quarterly/2025/02/iq2025q1-evtolevolution} are investing in low-altitude aviation and traffic management, while many countries are positioning LAE as a strategic sector beyond the digital economy and smart manufacturing\footnote{https://www.arabianbusiness.com/industries/transport/saudi-arabia-set-to-launch-autonomous-air-taxi-flights-this-year}. With enabling policies and continuous technological advances, LAE is poised to become a new infrastructure ecosystem that tightly integrates communication, sensing, computation, and airspace services.

Despite this momentum, the large-scale deployment of LAE still faces critical bottlenecks. First, terrestrial cellular networks, designed primarily for ground users, fall short in providing wide-area and three-dimensional coverage for aerial vehicles below 1km, especially in mountainous regions, offshore environments, and post-disaster scenarios where communication blind spots are common. Second, edge computing capabilities remain limited, as most services rely on remote cloud or ground data centers, which cannot satisfy the low-latency task offloading required by fast-moving drone swarms. Third, airspace regulation is still immature. That means the lack of unified unmanned traffic management (UTM)\footnote{https://eda.europa.eu/docs/default-source/documents/sc4-final-report-v1-0.pdf} and real-time monitoring makes it difficult to detect trajectory conflicts, boundary violations, or potential collisions in time, thereby raising safety concerns. These challenges collectively hinder LAE’s transition from pilot deployments to large-scale, routine operations, underscoring the need for a new type of aerial infrastructure that can simultaneously provide communication, computing, and supervisory functions.  
More critically, such fragmentation prevents the establishment of a unified airspace supervision and safety assurance layer, which is indispensable for scaling the LAE beyond pilot deployments toward routine and large-scale operations.

While low earth orbit (LEO) satellites have been regarded as a key enabler of LAE \cite{he2025satellite}, their limitations are becoming increasingly evident. LEO constellations excel at global connectivity and navigation, but their latency, though lower than geostationary earth orbit (GEO) satellites, remains inadequate for mission-critical tasks demanding millisecond-level responsiveness. Moreover, their emphasis on ``ubiquitous'' coverage comes at the expense of fine-grained, localized services in complex environments such as urban canyons, mountainous valleys, or disaster zones. Finally, LEO lacks intrinsic capabilities for low-altitude traffic monitoring and regulation, limiting its ability to support real-time trajectory prediction, collision avoidance, and safe swarm coordination.

\begin{figure*}[htp]
	\centering
	{\includegraphics[width=0.85\textwidth]{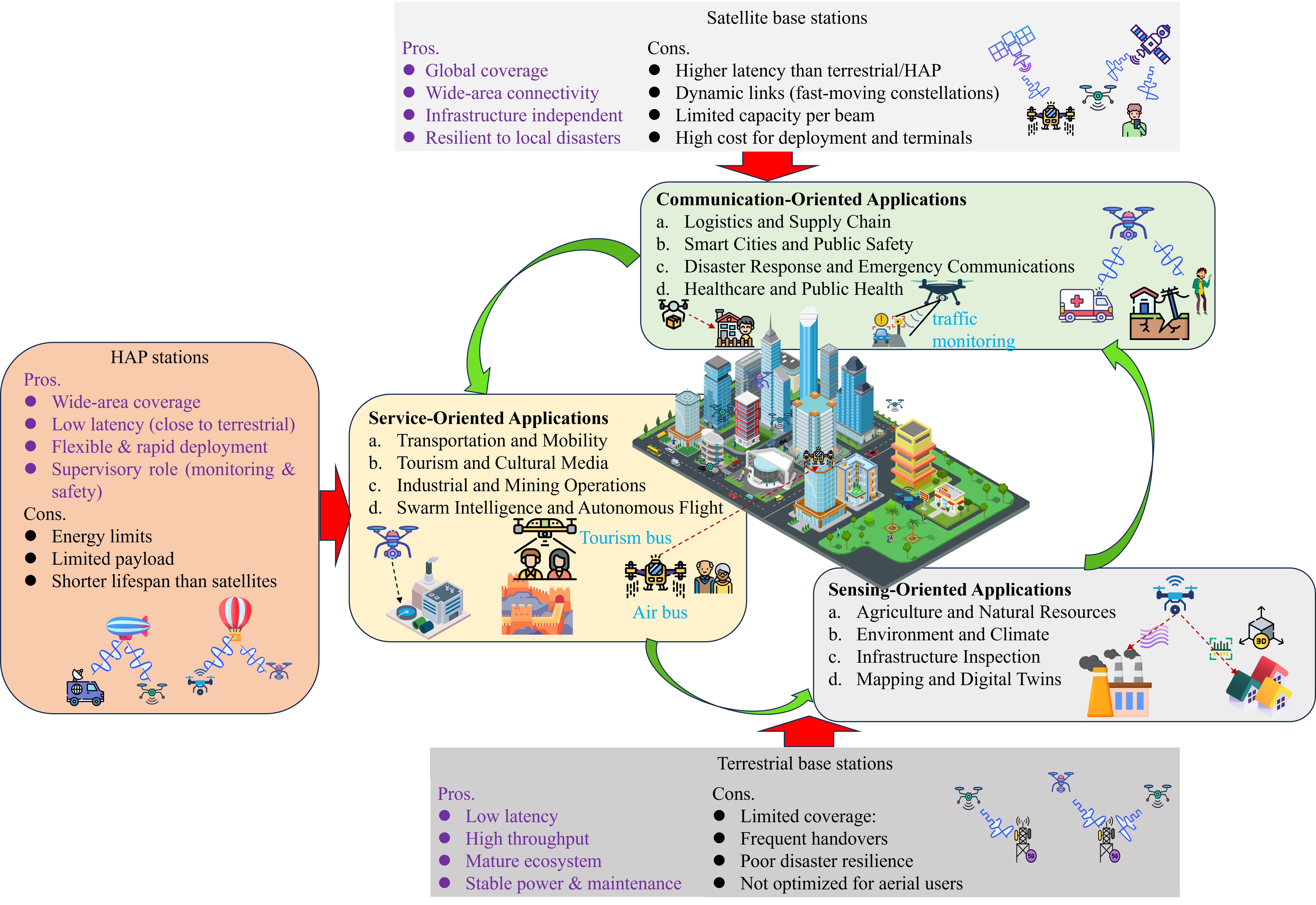}}
	\caption{Overview of LAE applications and supporting networks. The applications of the LAE can be categorized into three core domains: communication, sensing, and services. Communication focuses on addressing issues such as insufficient coverage, excessive latency, and discontinuous connectivity with providing information support. Sensing targets environmental monitoring and situational awareness, with an emphasis on data acquisition. Services highlight scheduling, regulation, and coordination, underscoring system-level integration and efficient operation. To enable these applications, a comprehensive air–space–ground network is leveraged, whose advantages and limitations are summarized.}
	\label{fig1}	
\end{figure*}

In this context, high-altitude platforms (HAPs) emerge not merely as additional access points, but as bridging nodes that integrate communication, computing, and airspace regulation into a coherent aerial control plane \cite{lou2023haps}. 
Operating in the stratosphere at around 20 km, HAPs strike a unique balance between wide-area coverage and low-latency responsiveness. They can deliver stable communication and sensing to both drones and ground users while acting as aerial ``super nodes'' that integrate edge computing, caching, and task scheduling. Furthermore, HAPs are naturally positioned to perform supervisory roles, monitoring unmanned aerial vehicle (UAV) trajectories, predicting risks, and issuing safety alerts, thereby complementing the capabilities of LEO satellites and terrestrial networks. Crucially, HAPs are unaffected by ground geography and can maintain continuous coverage in mountainous, offshore, or post-disaster environments, overcoming the blind spots of terrestrial networks. More details can be seen in Fig.~\ref{fig1}. Together with UAVs, LEO satellites, and ground/cloud infrastructures, HAPs can form a hierarchical global--regional--local architecture, laying the foundation for the scaled, safe, and intelligent development of the LAE.

Recognizing that the integration of HAPs into the LAE can effectively mitigate current deficiencies in communication coverage, computational offloading, and regulatory oversight, this article articulates an evolutionary trajectory for HAPs-enabled LAE. The envisioned role of HAPs can be delineated into five progressive stages:
\begin{itemize}
    \item \textbf{Foundational infrastructure platform}: HAPs serve as ``aerial bases'' that provide ubiquitous connectivity, edge computing, caching, and supervisory control, establishing the fundamental infrastructure required to support large-scale LAE operations.
    \item \textbf{UAV-centric service enabler}: Moving beyond basic infrastructure, HAPs evolve into dedicated service nodes for UAVs, offering reliable communication, high-precision sensing, navigation  integrity support, and mission offloading. This stage emphasizes direct enablement of UAV performance and autonomy, marking the shift from passive platform to active enabler.
    \item \textbf{Adaptive aerial--air--ground collaboration}: In complex or disrupted environments, HAPs cooperate with UAVs to provide localized coverage extension, environmental awareness, and resilient networking. This stage highlights context-aware and environment-driven collaboration, addressing scenarios where terrestrial and satellite systems are insufficient.
    \item \textbf{Swarm-scale orchestration hub}: As UAV operations scale, HAPs transform into orchestration centers for heterogeneous UAV swarms, enabling task allocation, cooperative sensing, collision avoidance, and large-scale trajectory optimization. Here, the focus shifts from single-UAV services to system-level coordination and swarm intelligence.
    \item \textbf{Cross-layer intelligent autonomy}: Ultimately, HAPs integrate with ground networks, cloud infrastructures, UAV collectives, and satellite constellations to form a space--air--ground--cloud closed-loop ecosystem. This holistic architecture enables real-time autonomy, adaptive optimization, and AI-driven orchestration across layers, thereby steering the LAE toward self-sustaining, intelligent, and resilient growth.
\end{itemize}

Through this staged progression, HAPs are expected to transcend their initial role as aerial relays and become a pivotal cross-layer, autonomy-driven infrastructure, underpinning the large-scale, intelligent, and sustainable evolution of the LAE. The remainder of this article further elaborates on these stages, highlighting the key enabling technologies, open research challenges, and prospective directions that will shape the future of HAPs-assisted LAE.

\section{Why HAPs, not (only) LEO?}

When envisioning the LAE, LEO satellites are often the first technology to be considered \cite{he2025satellite}. Indeed, dense LEO constellations provide global coverage, ensuring baseline connectivity across remote oceans, deserts, and international air corridors. Using a transportation analogy, LEO resembles a transnational expressway: highly effective for long-distance connectivity, but less adept at managing localized congestion or dynamic traffic control at specific intersections. What the LAE fundamentally requires, however, is a regional-scale traffic controller capable of fine-grained orchestration and safety management, this is the unique role of the HAPs. More specifically, the respective roles and differentiated impacts of LEO satellites and HAPs on the LAE can be contrasted as follows:
\begin{itemize}
    \item \textbf{Latency and control loops.}
Orbiting hundreds of kilometers above Earth, LEO satellites achieve lower latency than GEO systems, yet their tens-of-milliseconds delay remains insufficient for sub-second closed-loop operations such as UAV swarm collision avoidance, post-disaster emergency response, or urban airspace management. Additional impairments, including frequent handovers and Doppler shifts, further compromise stability. In contrast, HAPs stationed at approximately 20 km altitude offer millisecond-scale round-trip delays, approaching terrestrial performance and enabling rapid command–sense–feedback loops with enhanced resilience.

\item \textbf{Coverage and targeted services.}
Whereas LEO excels in ubiquitous macro coverage, HAPs specialize in localized, fine-grained services. A single HAPs can sustain “private-network-grade” support across 200–500 km regions, making it ideal for urban clusters, mountainous valleys, maritime corridors, major events, and post-disaster zones. Moreover, HAPs can be rapidly deployed within hours to provide on-demand emergency coverage, an agility that satellites inherently lack.

\item \textbf{Edge intelligence and energy efficiency.}
Low-altitude missions are typically data-intensive and computation-heavy, dominated by high-resolution imaging and real-time video streams. Offloading all tasks to the cloud or satellites is both delay- and energy-prohibitive, particularly for UAV with limited onboard resources. With greater energy reserves and payload capacity, HAPs can host AI accelerators and edge servers, effectively acting as “computing clouds in the sky.” By supporting task offloading, feature extraction, and in-situ data fusion, HAPs reduce UAV energy consumption while enhancing end-to-end efficiency and robustness.

\item \textbf{Airspace regulation and safety assurance.}
Unlike LEO satellites, which primarily function as global broadcasters, HAPs occupy a regulatory sweet spot, above civil aviation corridors yet below satellite orbits. From this vantage point, they can continuously monitor UAV trajectories, fuse heterogeneous sensing data, and predict boundary violations or collision risks. Coupled with ground-based UTM systems, HAPs serve as “intelligent traffic lights in the sky,” enforcing order and enhancing safety across congested low-altitude airspace.
\end{itemize}

Crucially, this does not diminish the role of LEO satellites. Instead, LEO and HAPs are complementary:
LEO provides the global backbone, delivering seamless, transnational connectivity;
HAPs deliver regional orchestration, edge intelligence, and airspace governance, ensuring localized efficiency and safety;
UAV and ground systems perform last-mile sensing, execution, and feedback.

A representative paradigm is a three-layer global–regional–local architecture. For example, in disaster scenarios, LEO satellites handle macro-level coordination and cross-domain aggregation, HAPs over the affected region provide real-time scheduling and edge computing, while UAV conduct street-level inspection, evidence collection, and material delivery. Only through multi-layered cooperation, with each component leveraging its strengths, can the LAE achieve ubiquitous coverage, effective regulation, and rapid responsiveness. Within this integrated ecosystem, HAPs emerge as the most application-proximate and indispensable enablers, transforming the global highways of LEO into regional-scale real-time governance and intelligent operations.

\section{HAPs Assistance in LAE}

\begin{figure}[htp]
	\centering
	{\includegraphics[width=0.45\textwidth]{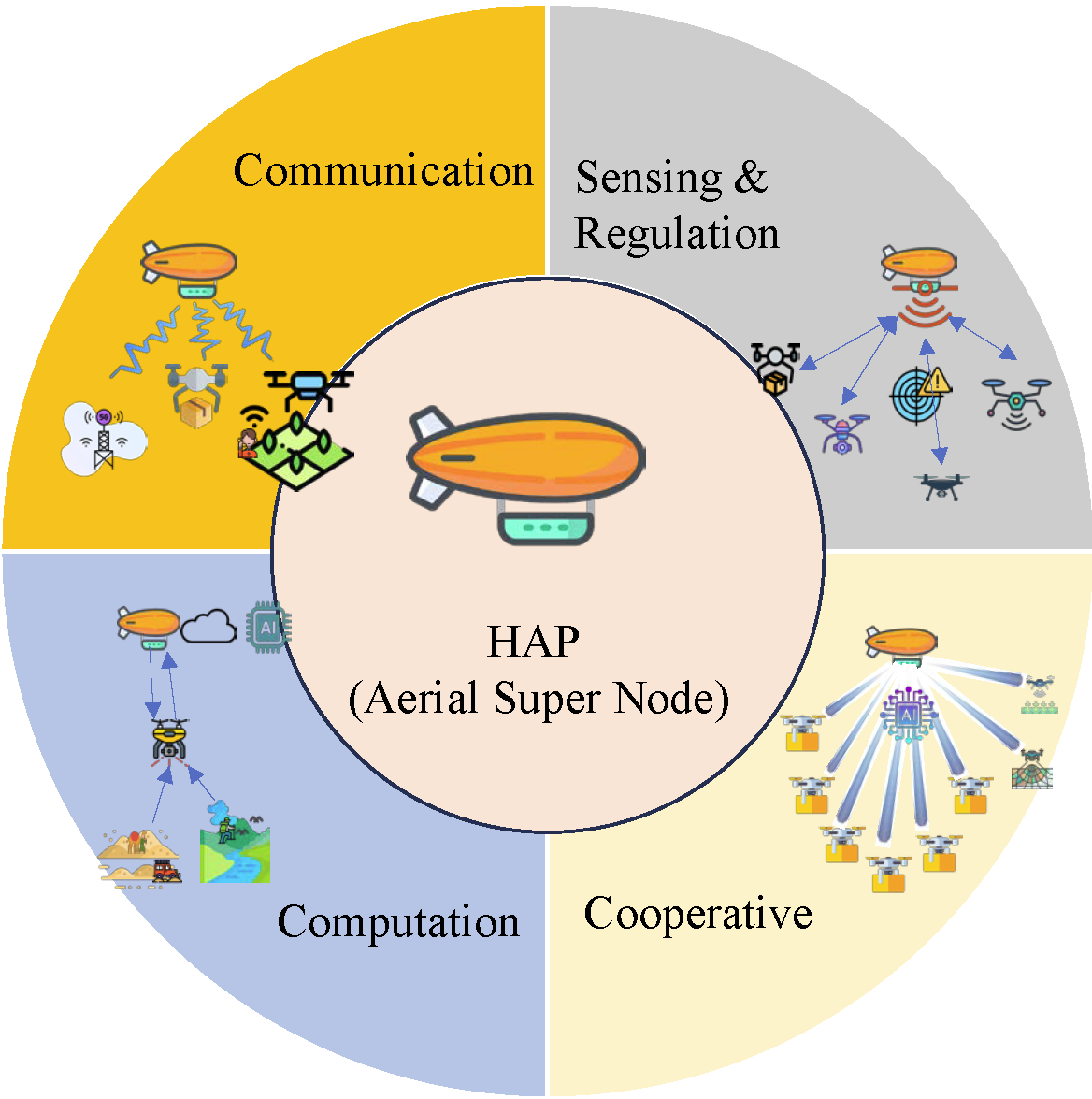}}
	\caption{HAPs as aerial super nodes in the LAE, enabling four key dimensions: Communication, Sensing $\&$ Regulation, Computation, and Cooperative Intelligence.}
	\label{fig2}	
\end{figure}

With the rapid expansion of the LAE, UAV are expected to perform increasingly diverse and mission-critical tasks in environments that are often dynamic, unstructured, and resource-constrained. However, the large-scale deployment of UAV is still constrained by bottlenecks in \emph{connectivity}, \emph{regulation}, \emph{computational capability}, and \emph{swarm-level coordination}. HAPs, positioned in the stratosphere at approximately 20 km, offer a unique vantage point and resource pool to address these challenges. Specifically, the assistance provided by HAPs in LAE can be categorized into four complementary dimensions: communication, sensing and regulation, computation offloading, and cooperative intelligence, as shown is Fig.\ref{fig2}.  
\subsection{Communication Assistance: Stable and Wide-Area Aerial Connectivity}

In the LAE, UAV operations span diverse and complex environments, from emergency logistics over mountainous terrain to inspection tasks in dense urban corridors. However, traditional terrestrial cellular networks are constrained by limited coverage radius and vertical beam patterns, leading to frequent handovers and service interruptions. These factors create critical bottlenecks for mission continuity and real-time command reliability.
HAPs offer a macro-scale communication anchor that complements and enhances ground infrastructure. Functioning as aerial macro base stations, HAPs extend coverage across hundreds of kilometers and significantly reduce handover frequency. Moreover, they can be integrated with terrestrial networks to form dual-path protected communication links, where the HAPs link seamlessly takes over when the ground link experiences blockage or interference. For UAV swarms, HAPs also enable centralized spectrum coordination and routing management, effectively mitigating intra-swarm interference and supporting scalable, cooperative connectivity.

\begin{figure*}[htp]
	\centering        
	\subfloat[]{
    \includegraphics[width=0.45\textwidth]{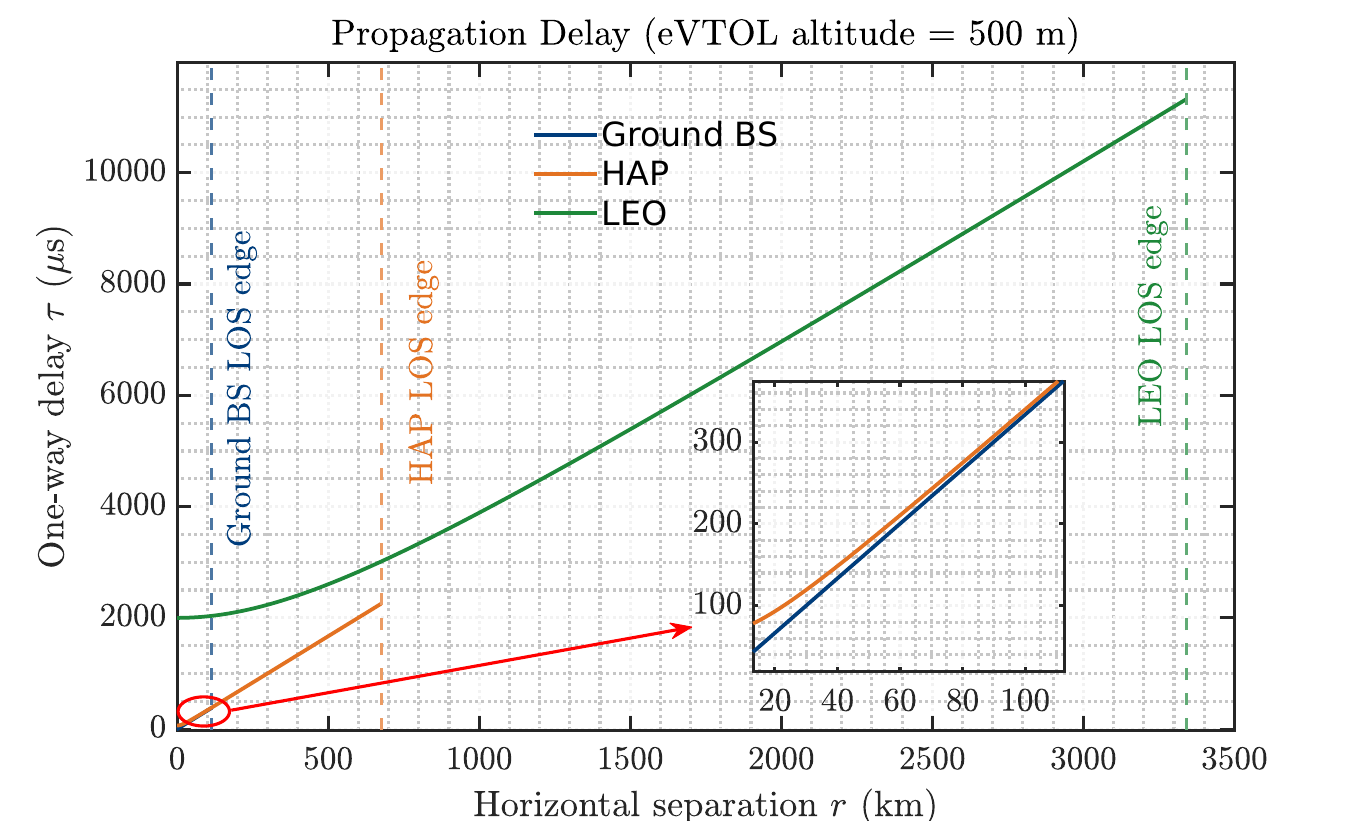}\label{fig_11}}
	\subfloat[]{
		\includegraphics[width=0.45\textwidth]{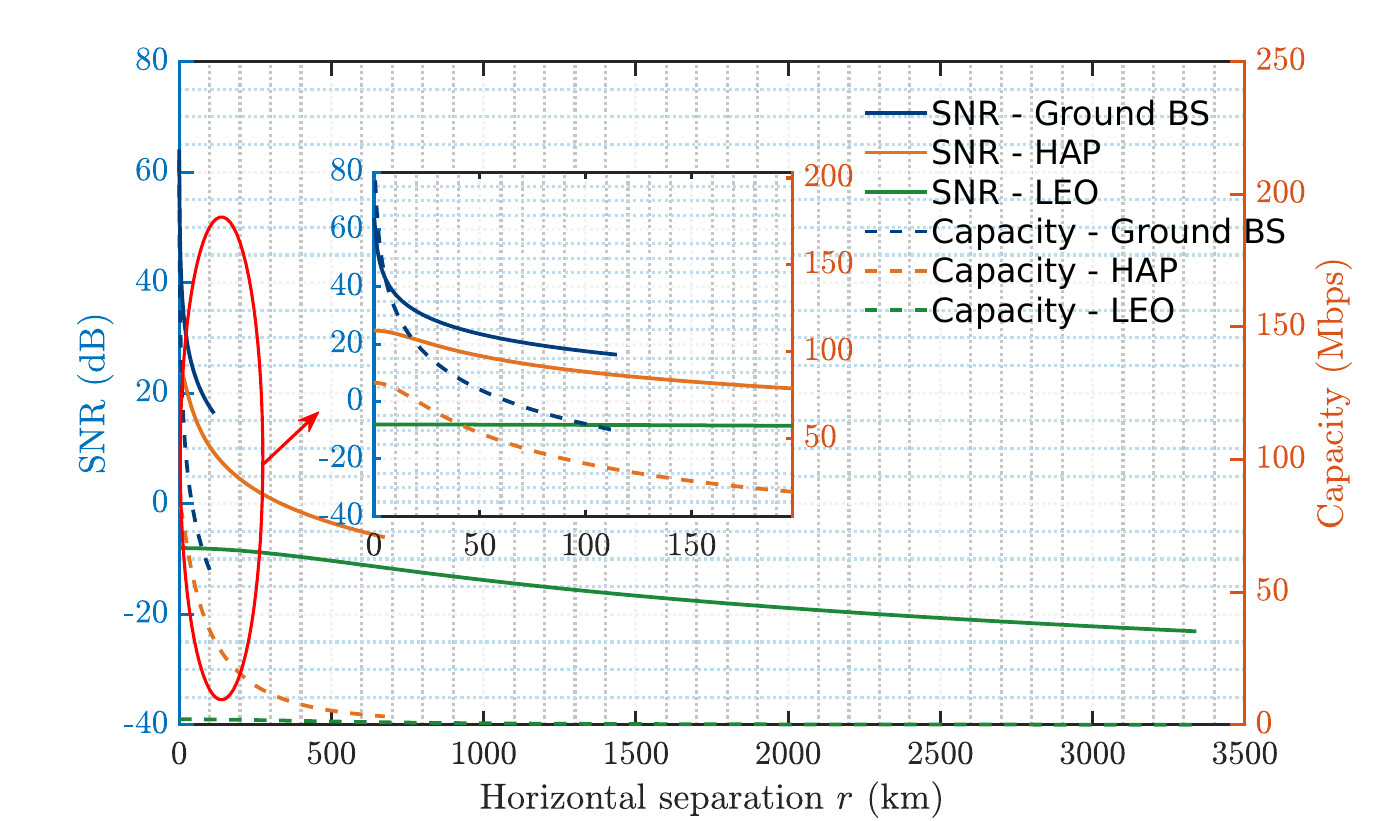}\label{fig_12}}  
	\caption{Comparison of communication characteristics across different infrastructures for eVTOL connectivity. (a) One-way propagation delay versus horizontal separation distance, (b) Corresponding SNR and achievable capacity.}
    \label{fig_1}
\end{figure*}

As illustrated in Fig.~\ref{fig_1}, Ground BSs ensure low delay and strong signal to noise ratio (SNR) only within a short service radius, while LEO satellites offer global access but at the cost of higher propagation delay and signal attenuation. HAPs achieve a balanced performance, providing both low-latency and wide-area coverage, along with more stable SNR over long distances. This makes HAPs particularly well-suited as the communication backbone for large-scale, mobile, and safety-critical UAV/eVTOL swarm operations in the LAE.

\subsection{Sensing, Regulation, and Navigation Integrity Assistance: Airspace Safety and Integrated Sensing--Communication}

The large-scale expansion of the LAE requires not only ubiquitous communications but also effective regulation and safety assurance across densely populated and highly dynamic low-altitude airspace. Leveraging its high-altitude vantage point, the HAPs can continuously monitor UAV trajectories and recognize abnormal behaviors across wide areas. By incorporating onboard AI models, the HAPs can predict potential collision risks and issue graded early warnings, effectively functioning as an intelligent radar for airspace management.

Beyond monitoring, the HAPs can employ integrated sensing–communication (ISAC) waveforms~\cite{TangYu2025Cooperative} to simultaneously support UAV navigation, velocity estimation, and positioning while maintaining data transmission, thus reducing the need for dedicated sensing payloads. In complex environments such as canyons or dense urban areas, the HAPs can cooperate with UAVs or reconfigurable intelligent surfaces (RIS)~\cite{AhmedSoofi2025Towarda} to mitigate blind spots and improve coverage continuity. In this sense, the HAPs serves not only as the communication backbone but also as the air traffic coordinator and safety guardian of the LAE. In addition to providing wide-area situational awareness, persistent HAPs-based sensing also enables {navigation integrity support} for UAV and eVTOL operations. By maintaining high-probability line-of-sight links and continuous trajectory supervision, HAPs can complement the Global Navigation Satellite System (GNSS) in challenging environments such as urban canyons or obstructed regions, thereby enhancing trajectory reliability and airspace trust. 

It is worth noting that such navigation assistance is realized at the {system and regulation level}, rather than through dedicated onboard positioning algorithms. Instead of pursuing standalone localization accuracy, HAPs contribute to navigation integrity by jointly leveraging sensing, communication, and regulatory supervision to ensure trustworthy and conflict-free aerial operations.

\begin{figure*}[t]
    \centering
    \subfloat[]
    {\includegraphics[width=0.23\textwidth]{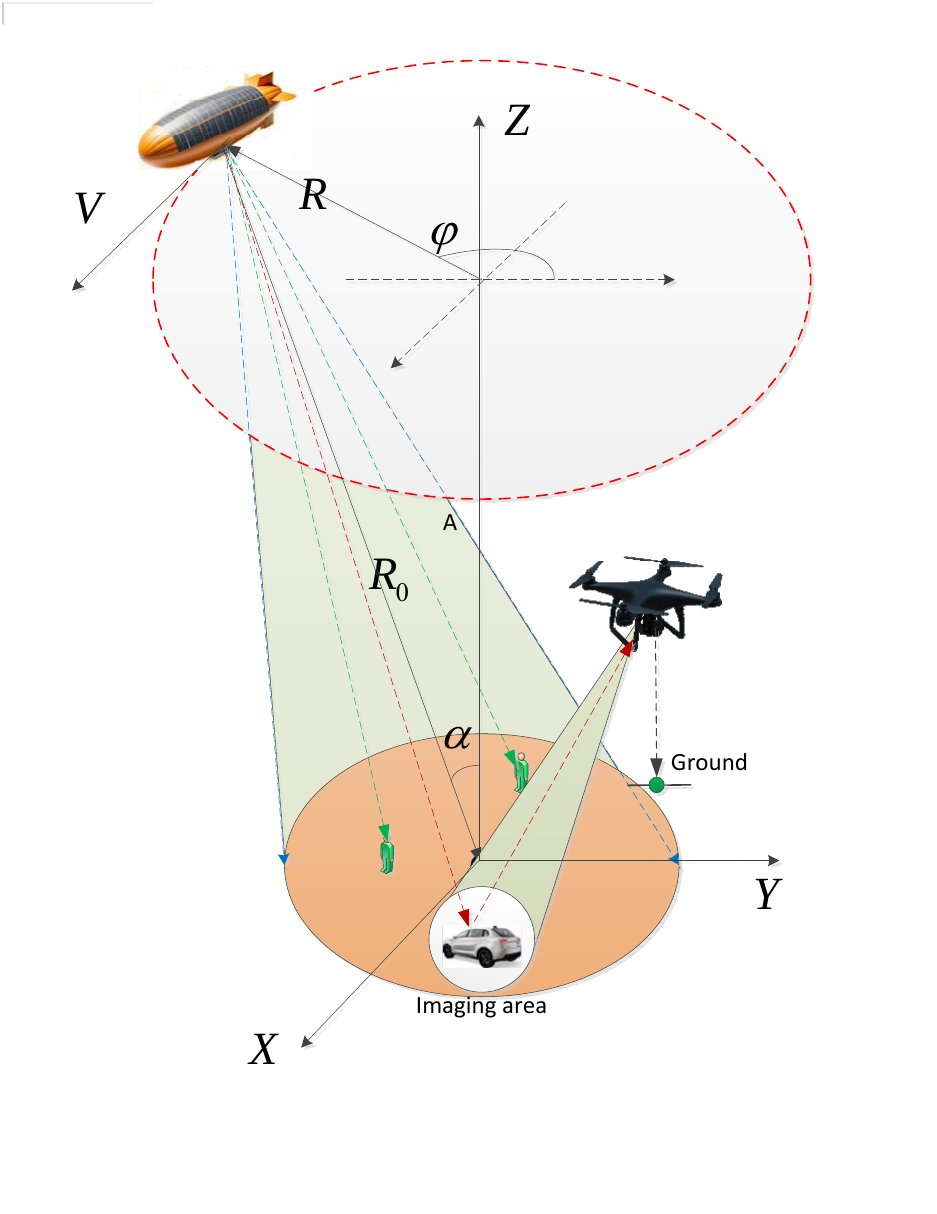}
    \label{fig:isarac_scenario}}
    \hfill
    \subfloat[]
    {\includegraphics[width=0.37\textwidth]{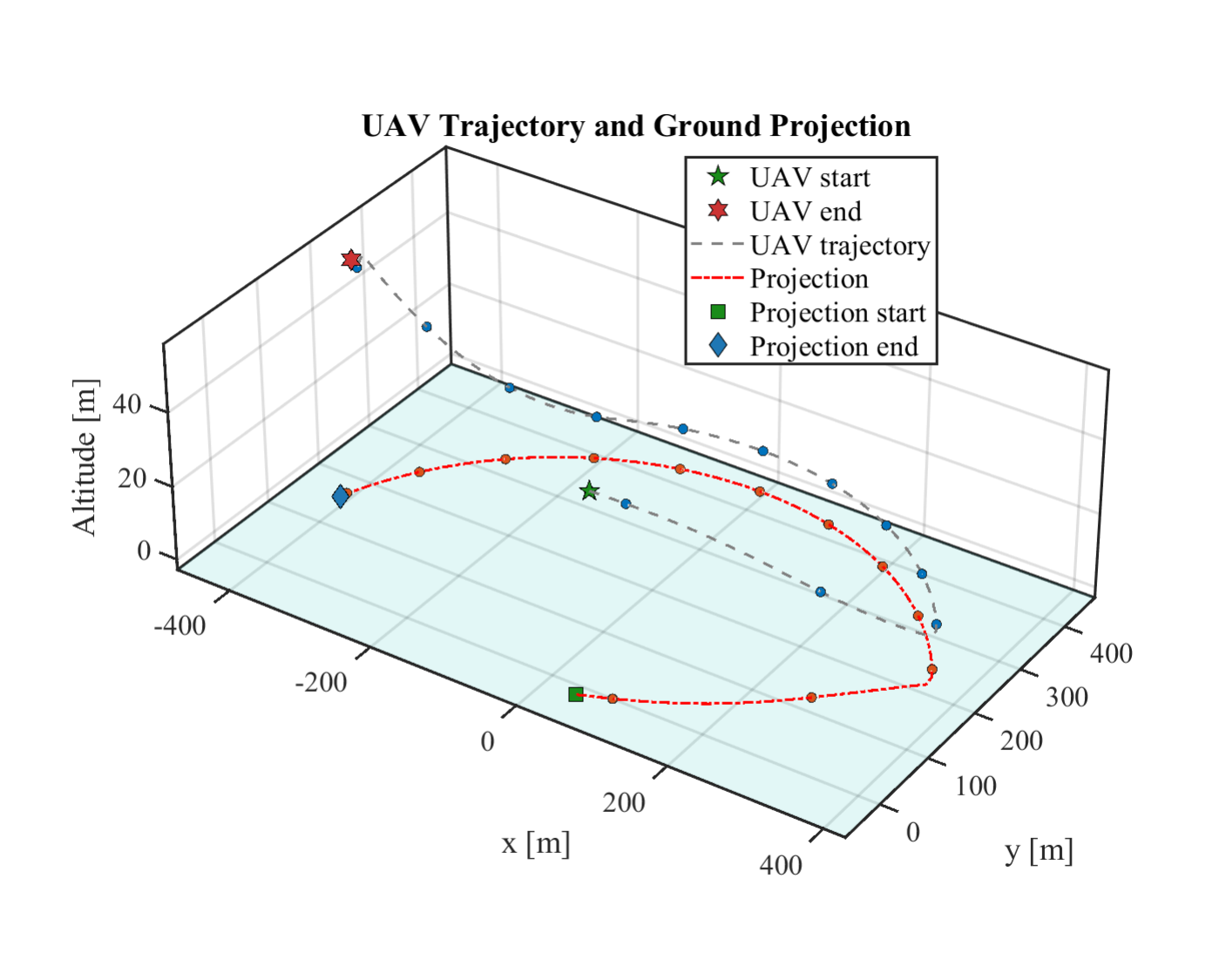}
    \label{fig:isarac_traj}}
    \hfill
    \subfloat[]
    {\includegraphics[width=0.37\textwidth]{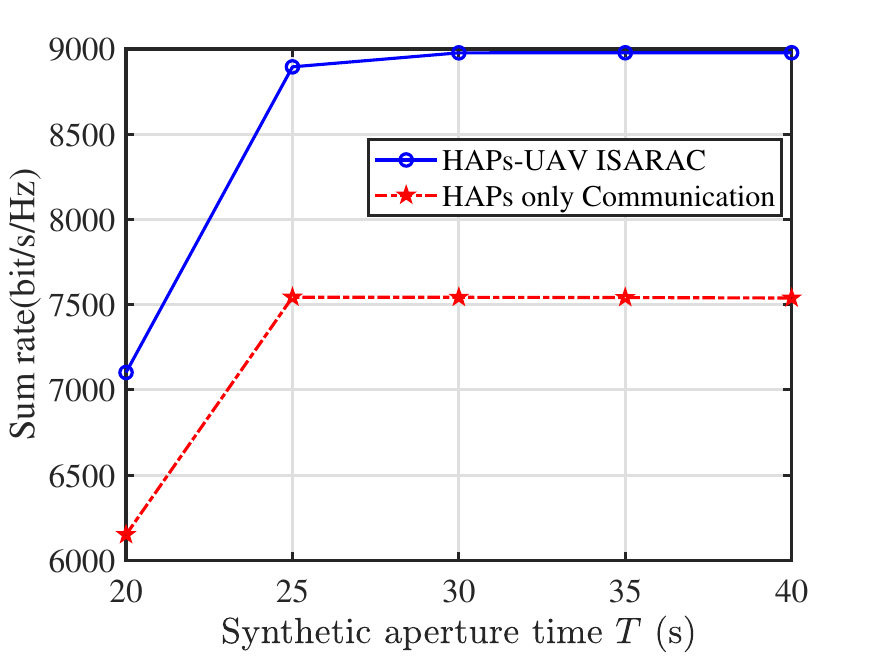}
    \label{fig:isarac_rate}}
\caption{Representative simulation results showing (a) the system model of a cooperative HAPs–UAV bistatic ISARAC scenario, (b) the optimized UAV trajectory that enhances synthetic aperture diversity while maintaining a reliable downlink for ISAC illumination from the HAPs, and (c) the ISAC performance trend illustrating balanced throughput and sensing robustness under shared energy constraints.}
    \label{fig:isarac_results}
\end{figure*}

To provide a more tangible view of this concept, Fig.~\ref{fig:isarac_results} presents representative simulation results from our recent HAPs–UAV integrated sensing and communication (ISARAC) study~\cite{Huang2025_ISARAC}.
Figure~\ref{fig:isarac_traj} shows the optimized UAV trajectory under a joint sensing–communication configuration. In this setup, the HAPs transmits an ISAC waveform that simultaneously enables downlink communication and synthetic aperture illumination, while the UAV acts as a passive receiver collecting echoes for bistatic SAR processing. The optimized flight path provides sufficient aperture diversity for imaging without compromising the quality of the downlink channel.
Figure~\ref{fig:isarac_rate} illustrates the overall system performance trend, highlighting how integrated resource allocation affects both throughput and sensing reliability under shared power and energy constraints. Although the exact numerical values depend on parameter settings, the observed trends demonstrate that the cooperative operation between HAPs and UAVs achieves balanced resource utilization, maintaining robust connectivity while enabling high-fidelity sensing.
Together, these insights qualitatively validate the feasibility and efficiency of HAPs-assisted ISAC architectures in enhancing situational awareness, airspace safety, and wide-area connectivity within the LAE~\cite{Huang2025_ISARAC}.

\subsection{Computation and Offloading Assistance: Aerial Cloud for UAV Computing}

Many UAV applications, such as high-definition video streaming, target detection, and remote imaging, require intensive computation. However, UAVs are inherently constrained by limited onboard processing capability and battery capacity. This creates a fundamental tension between the sensing demands of the mission and the platform's available computational resources.
To address this limitation, HAPs can function as an “aerial cloud data center,” providing computation offloading services over a wide coverage area. In environments where terrestrial connectivity is weak or unavailable (e.g., mountainous regions or post-disaster zones), the UAV can adopt a “heavy sensing, light reporting” strategy: raw data is captured onboard but offloaded to the HAPs for feature extraction, inference, and recognition, while only essential results are fed back. This approach ensures task continuity while avoiding excessive onboard energy consumption.

In addition, HAPs can jointly optimize computation–caching–communication scheduling \cite{JiaWu2023HierarchicalAerialComputing}. During low-traffic periods, data or AI model pre-processing can be conducted proactively at the HAPs, whereas real-time computation tasks are prioritized during peak mission windows. This coordinated architecture ensures the UAV maintains reliable computational support even under tight energy budgets or heavy sensing loads.

\begin{figure*}[htp]
	\centering        
	\subfloat[]{
    \includegraphics[width=0.45\textwidth]{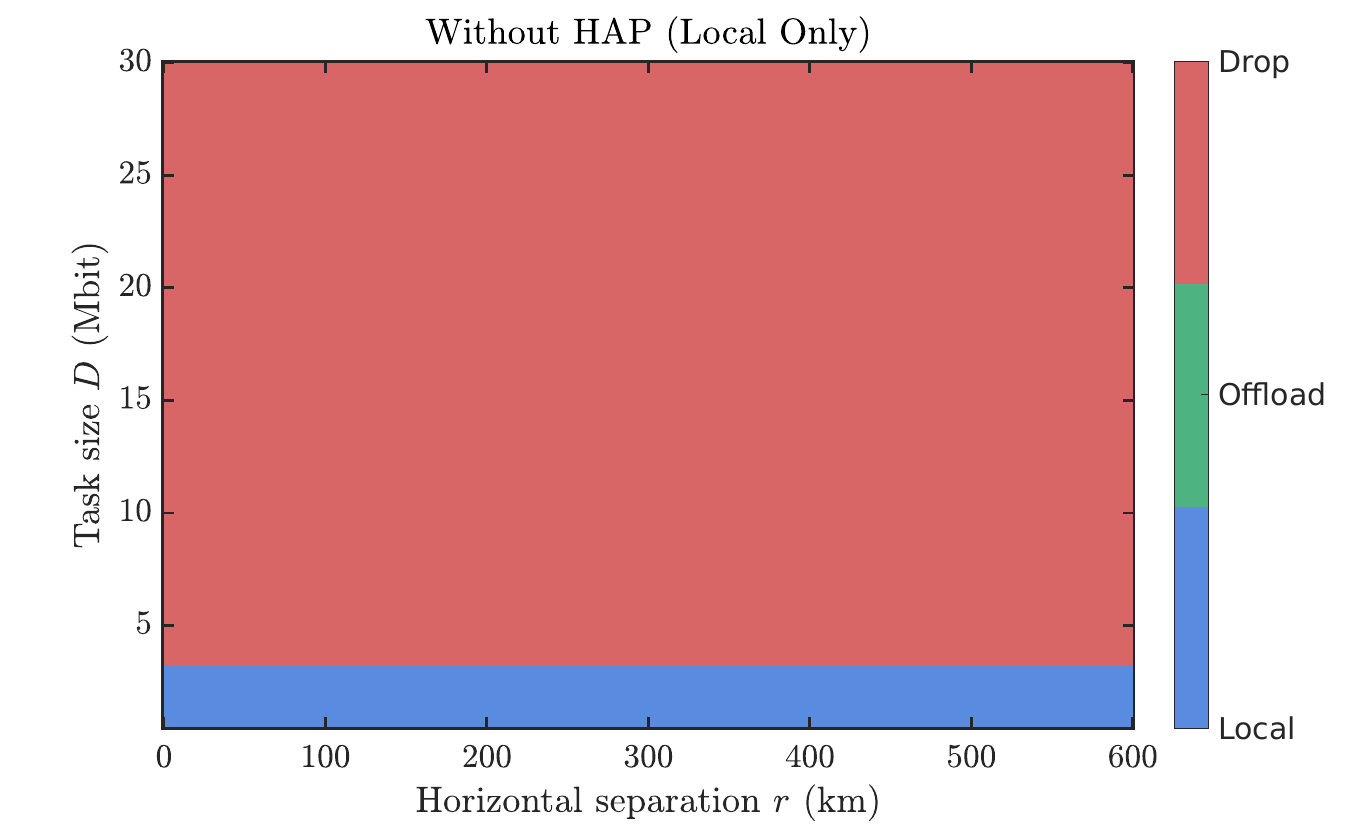}\label{fig_51}}
	\subfloat[]{
		\includegraphics[width=0.45\textwidth]{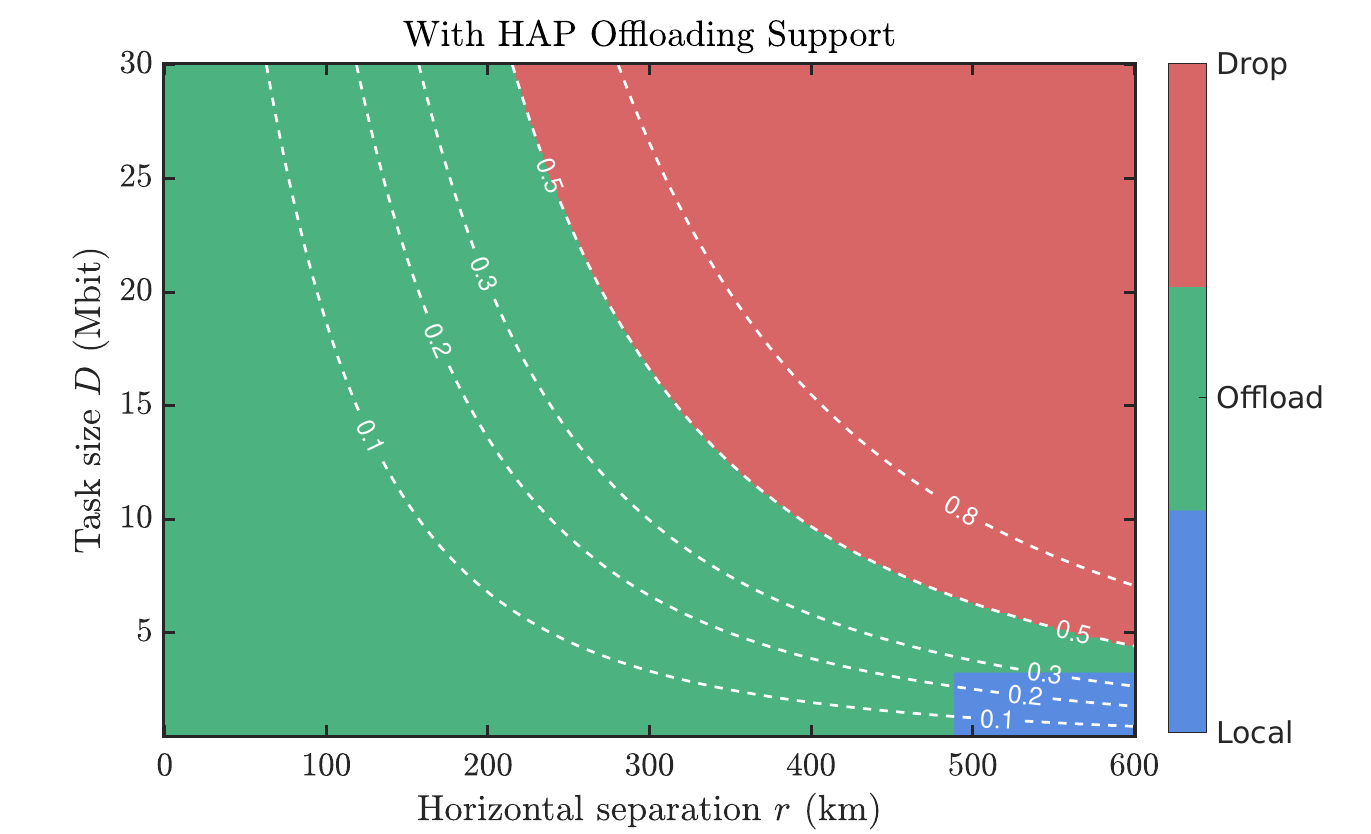}\label{fig_52}}  
	\caption{Impact of HAPs-enabled computation offloading on UAV task executability.
(a) Without HAPs. (b) With HAPs offloading support
.}
    \label{fig_5}
\end{figure*}

As shown in Fig.~\ref{fig_5}, relying solely on local computation severely restricts the UAV’s computable task region. Without HAPs support (Fig.~\ref{fig_51}), only tasks with relatively small data sizes can be completed within the latency bound $T_{\max}$, while a large portion of tasks become infeasible (red region), reflecting strict limits in onboard computing and battery resources.
When HAPs offloading is enabled (Fig.~\ref{fig_52}), a substantial portion of these previously infeasible tasks becomes executable (green region). The feasible region expands simultaneously toward larger task sizes and greater UAV–HAPs separation distances, highlighting the role of HAPs as an aerial cloud extending computation capability across wide-area airspace. The white dashed contours denote the end-to-end offloading delay, showing that offloading remains within the latency constraint over a broad range. Consequently, HAPs offloading is well-suited for time-sensitive UAV missions such as disaster response, environmental monitoring, and search-and-rescue.
In essence, HAPs transforms the UAV from a resource-limited sensing node into a computation-augmented intelligent platform, enabling more complex, data-intensive, and long-duration aerial missions.

\subsection{Cooperative Intelligence Assistance: Swarm Intelligence and Closed-Loop Autonomy}

As UAV deployments scale up, single-drone operations are evolving into collaborative swarm missions. In such scenarios, HAPs functions as an aerial coordination tower, responsible for assigning inspection regions, balancing mission loads, and coordinating intra-swarm routing and spectrum access so as to prevent link congestion and task conflict.  
Beyond task allocation and control signaling, HAPs also facilitate multi-modal sensing fusion, integrating heterogeneous measurements such as radar echoes, mmWave returns, and optical imagery collected by UAV swarms. Performing fusion and consistency filtering at the HAPs ensures robust perception, particularly in complex or dynamic environments. At a higher architectural level, UAVs, HAPs, and terrestrial/cloud infrastructures form a three-tier edge–air–cloud autonomy hierarchy, where UAVs perform fast front-end detection, HAPs execute regional reasoning and mission-level decision-making, and cloud servers handle global optimization and long-term strategy planning. This edge–air–cloud closed-loop autonomy framework enables UAV swarms to complete the full cycle of detection–sensing–planning–control within second-level latency, thereby achieving true swarm intelligence.

As shown in Fig. 6, the total network throughput depends critically on how co-channel interference is managed.
In the No coordination baseline, UAVs randomly select channels, causing severe mutual interference as swarm size increases. The Power-only strategy reduces interference by lowering transmit power to meet SNR targets, but this results in overly conservative spectrum usage, offering only marginal throughput gains.
In contrast, the proposed HAPs spectrum coordination strategy performs joint link-quality estimation, channel assignment, and bandwidth allocation at the HAPs level. By exploiting UAV spatial distribution and traffic load, the HAPs minimizes co-channel interference and reallocates bandwidth proportionally to achievable link rates. As a result, the network maintains scalable throughput growth even at large swarm sizes, significantly outperforming both baselines.

\begin{figure}[htp]
	\centering        
	\subfloat[]{
		\includegraphics[width=0.45\textwidth]{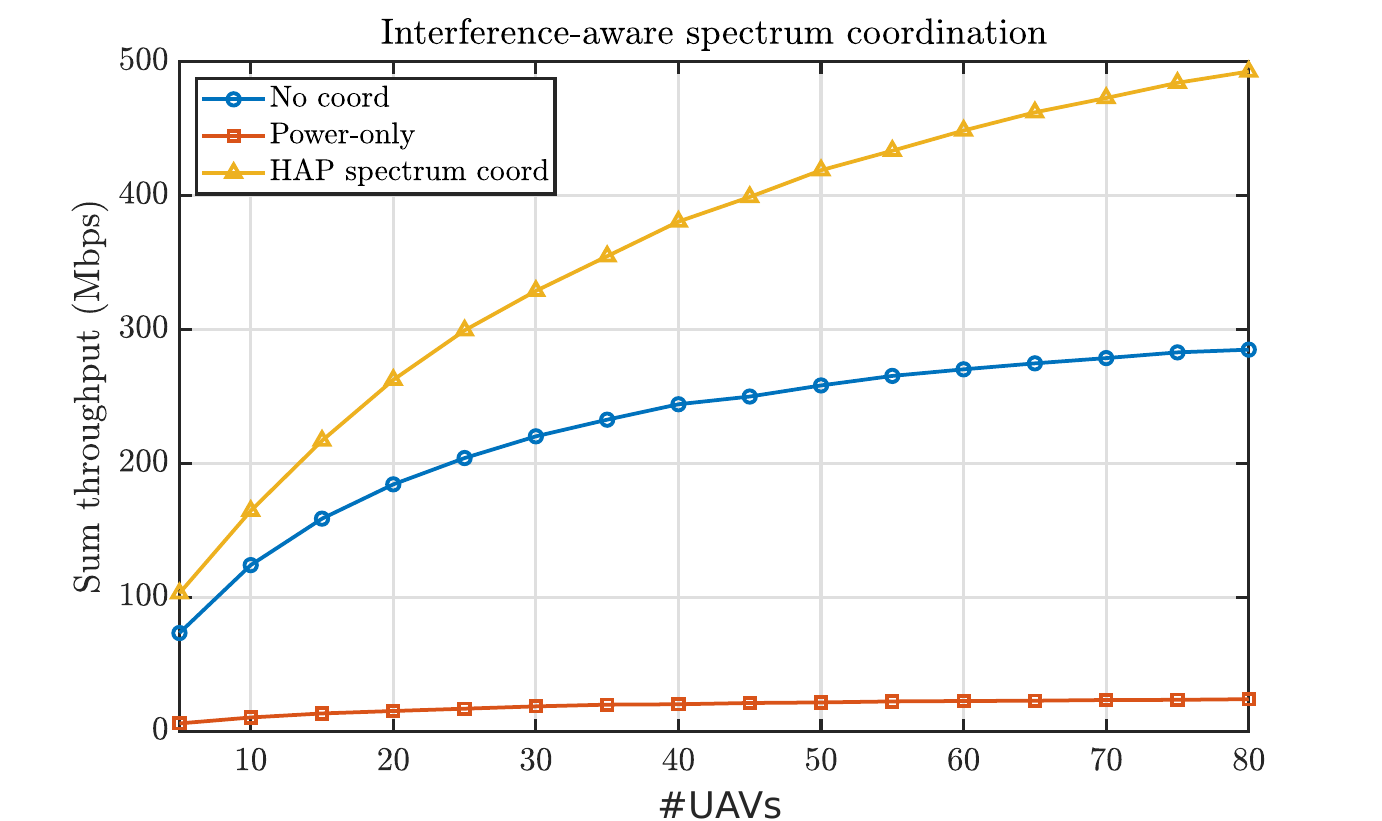}\label{fig_52}}  
	\caption{Sum throughput performance under different interference management strategies.}
    \label{fig_5}
\end{figure}

\subsection{Summary}

These four modules directly address the most urgent demands of the LAE, namely reliable communication, secure regulation, computational empowerment, and swarm intelligence. Importantly, they are not isolated but mutually reinforcing: communication ensures stable data exchange, sensing and regulation provide safety guarantees, computation offloading enhances efficiency and endurance, while cooperative intelligence enables scalable swarm-level autonomy. Collectively, these capabilities enable HAPs to function as aerial control anchors that jointly support connectivity, intelligence, trusted navigation and airspace governance for large-scale and safety-sensitive low-altitude operations. As a result, the integration of HAPs establishes an intelligent aerial substrate that does not merely supplement terrestrial networks, but fundamentally reshapes the operational resilience, scalability, and intelligence of the LAE.

\section{Evolutionary Roadmap of HAPs in the LAE}

The development of the LAE is not a sudden leap, but rather a process of gradual accumulation and stepwise transformation. Within this evolving ecosystem, UAV can be conceptualized as playing two complementary roles. On the one hand, they operate as service providers, delivering communication, sensing, and other functionalities to ground users and IoT devices. On the other hand, they also act as service consumers, depending on ground base stations, HAPs, or LEO satellites for navigation, connectivity, and control. Existing research has primarily focused on the former role \cite{KananiOmidi2025OptimizingNetworkPerformance}, whereas the latter remains largely unexplored \cite{yan2025hierarchical}, thereby highlighting a critical research gap.

Against this backdrop, HAPs are expected to undergo a paradigm shift: moving beyond their traditional role as infrastructure providers to emerge as intelligent decision-makers that orchestrate UAV operations and ensure safe, reliable, and adaptive LAE services. This transformation can be systematically delineated into five progressive stages, outlining an evolutionary roadmap that spans from basic platform functions to full autonomy.

\subsection{Stage I: Infrastructure Layer (Aerial Base)}
Driven by the lack of unified coverage and fragmented terrestrial infrastructures, the initial stage positions HAPs as stable aerial bases. They serve as wide-area access points for UAV and ground users, supporting computation offloading and caching distribution, while establishing unified spectrum regulation and airspace monitoring capabilities. At this stage, HAPs function as the “6G mega-towers in the sky”, consolidating fragmented communication and computing resources into an integrated aerial network.

\subsection{Stage II: Super Backstage for UAV}
As UAV missions become increasingly heterogeneous and computation-intensive, HAPs upgrade their role into a super backstage. Beyond providing stable uplink and downlink connectivity, HAPs employ ISAC waveforms to enable simultaneous navigation and sensing. UAV can also offload computation-intensive imaging and detection tasks to HAPs for rapid processing. In this stage, HAPs evolve beyond being mere “base stations” and act as \emph{super servers} for UAV.

\subsection{Stage III: Frontline Support for Ground Users}
Motivated by the urgent demand for resilient connectivity in disaster and obstructed environments, HAPs begin to directly serve ground users. They can be rapidly deployed to restore communications in disaster areas. In valleys or dense urban environments, they ensure reliable signal delivery. At this stage, HAPs transition from \emph{backstage support} to \emph{frontline support}, becoming indispensable for the functioning of ground society.

\subsection{Stage IV: Swarm-Scale Coordination}
With the exponential growth of UAV fleets and the limitations of individual scheduling, HAPs emerge as aerial coordination towers for large swarms. They centrally orchestrate spectrum and routing to prevent congestion and interference. Leveraging global situational awareness, HAPs dynamically allocate mission areas, ensuring efficient and orderly UAV collaboration. Here, HAPs transcend the role of communication nodes to become the \emph{neural centers} of UAV swarm intelligence.

\subsection{Stage V: Full Autonomy with Edge–Air–Cloud Synergy}
Driven by the need for real-time intelligence and seamless multi-layer integration, HAPs integrate with ground edge nodes and cloud platforms to form a three-tier computational synergy architecture. UAV perform real-time detection, HAPs conduct regional reasoning and mission-level decision-making, while the cloud executes global optimization. Through this closed loop, the entire system completes the \emph{sense–decide–act} cycle within second-level latency, achieving true high-level autonomy. This signifies that HAPs are no longer merely aerial \emph{support platforms}, but the \emph{intelligent brain} of the LAE ecosystem.

The evolutionary roadmap of HAPs in the LAE illustrates a gradual yet transformative trajectory: from filling infrastructure gaps and enhancing UAV missions, to enabling resilient ground services, orchestrating swarm-scale operations, and ultimately achieving full autonomy through edge–air–cloud synergy. Each stage reflects not only a response to pressing technical challenges but also a step toward a broader vision of intelligent and sustainable low-altitude ecosystems. By reshaping how UAV and ground systems interact, HAPs are set to become the intelligent backbone of the LAE, guiding both technological innovation and policy directions toward a safer, more resilient, and economically vibrant future.

\section{Future Research Directions}

Although HAPs have already demonstrated unique potential in the LAE, this remains a rapidly evolving research area. The future development of HAPs will not be limited to acting as “communication relays” or “aerial bases,” but will advance toward higher levels of intelligence, coordination, and sustainability. The following four directions outline the core research pathways for the next phase.  

\subsection*{1) AI-Driven HAPs Intelligent Scheduling}  
Future HAPs will no longer rely on static deployment but will possess capabilities for autonomous trajectory prediction and dynamic scheduling. By leveraging deep reinforcement learning (DRL) and large-model-based scenario forecasting, HAPs can adaptively adjust altitude, position, and beam distribution according to UAV activity density, communication demand distribution, and environmental dynamics, thereby realizing on-demand coverage. This intelligent scheduling transforms HAPs into truly {adaptive network nodes}, rather than passive base stations.  

\subsection*{2) HAPs–LEO Cooperative NTN}  
The cooperation between HAPs and LEO satellites represents a key trend for future Non-Terrestrial Networks (NTNs). LEO satellites provide global coverage, while HAPs deliver fine-grained regional services. Their integration enables the construction of a holistic space–air network that combines global and regional advantages. In this architecture, LEO satellites handle intercontinental transmission and large-scale broadcasting, while HAPs provide localized, high-throughput, low-latency management. For applications such as cross-border logistics, maritime surveillance, and post-disaster emergency response, this complementary NTN paradigm will be essential for achieving truly seamless connectivity.  

\subsection*{3) Integrated Airspace–Spectrum–Computation Management}  
With the exponential growth of UAV deployments, future challenges will extend beyond communication links to encompass the integrated management of {airspace, spectrum, and computation}. HAPs are expected to become pivotal nodes for UTM: they can allocate airspace to avoid trajectory conflicts, while dynamically coordinating spectrum and computational resources to ensure efficient swarm-level task execution under resource constraints. Such cross-dimensional orchestration positions HAPs as the {grand coordinators} of the LAE.  

\subsection*{4) Green and Sustainable HAPs}  
For HAPs to support large-scale applications, issues of {energy and cost} must be addressed. Future research will focus on solar-powered designs, lightweight balloons, and cost-effective aerostats to achieve long-endurance station-keeping and reusable deployment. Combined with intelligent energy management and eco-friendly materials, HAPs will evolve toward {low-carbon and sustainable platforms}. This direction not only aligns with global trends in green communications but also significantly lowers the threshold for scaling the LAE.

In summary, the future evolution of HAPs will be characterized by four intertwined trajectories: {intelligence} (AI-driven autonomy), {globalization} (HAPs–LEO cooperation), {integration} (joint airspace–spectrum–computation management), and {sustainability} (green energy and low-cost platforms). These research directions will determine whether HAPs can transition from experimental testbeds into the {core infrastructure} that sustains the LAE at scale.  

\section{Conclusion}

The LAE is emerging as the next technological and industrial frontier. To truly unlock its potential, however, a stable and intelligent aerial infrastructure is indispensable. In this process, HAPs play a dual role, serving both as the {aerial base} and as the {coordination tower} for airspace governance.
Compared with LEO satellites, HAPs are much closer to users and thus capable of providing millisecond-level real-time communication. Leveraging their stratospheric vantage point, HAPs also assume the functions of airspace supervision and risk prediction. Equipped with onboard computing and caching resources, they further enable intelligent processing and task offloading for UAVs and ground users. Through the integration of communication, computing, and regulation, HAPs further enable navigation integrity and airspace trust as emergent system-level properties, paving the way toward safe, scalable, and intelligent deployments of the LAE.

At the application level, the synergy between HAPs and UAVs unleashes tremendous potential. HAPs can not only provide communication assurance and sensing offloading for UAV swarms but also be rapidly deployed in disaster-stricken areas to restore connectivity, perform sensing imaging, and deliver frontline support for rescue and logistics. Beyond this, HAPs complement LEO satellites, terrestrial cellular systems, and cloud computing resources to form a three-tier global–regional–local architecture: LEO satellites ensure global connectivity, HAPs enable regional fine-grained management, and UAVs execute localized detection and operations.

Looking ahead, HAPs are poised to become pivotal nodes that connect air traffic management, intelligent logistics, and emergency response. They will not only enhance the scalability of LAE operations but also drive the ecosystem toward high autonomy and sustainable growth. From functioning as an {aerial base} to evolving into an {intelligent brain}, HAPs are reshaping the foundational logic of the LAE and accelerating its transition from exploration to large-scale, autonomous deployment.

\bibliographystyle{IEEEtran}
\bibliography{references}




\end{document}